*Coherent x-ray radiation induced by high-current breakdown on a ferrite surface*

Ivan N. Tilikin, Sergey Yu. Savinov, Nikolai V. Pestovskii, Sergey A. Pikuz, Sergey N. Tskhai, and Tatiana A. Shelkovenko

P.N. Lebedev Physical Institute of RAS, 119991 Moscow, 53 Leninskii prosp.

*We for the first time observe that at the initial stage of a high-current discharge, a low-divergence short (< 2 ns) electromagnetic pulse is formed over a ferrite surface. The 50% part of this pulse lies in the region of fairly hard x-ray radiation (hν > 1 keV) with the energy ~0.6 mJ and the average power 0.3 MW. The radiation propagates parallel to the surface in the anode direction with the angle divergence < 2°. The high directionality of the radiation in absence of the aperture-limiting devices for the radiation beam and the quadratic dependence of the spatial radiation energy flux density on the active part of the ferrite prism points to the coherent nature of the observed radiation. A possible generation mechanism of the radiation is proposed. It is based on the short-lived magnetization of the unit areas on the ferrite surface by a high-power electromagnetic pulse and subsequent coherent interference of the unit waves irradiated by these areas.*

In the studies of vacuum ultraviolet (VUV) plasma radiation in a high-current discharge on a ferrite surface [1,2], a short ($\tau$ < 2 ns) directional pulse of electromagnetic radiation was detected at the initial stage of the discharge. The 50% energy part of this radiation is fairly hard - the corresponding energies of photons are higher than 1 keV. In this paper, we report the first results on the study of this radiation and provide their possible interpretation. We show in the present work that the radiation is has the low angle divergence (< 2°) and its high radiant intensity can be caused by the coherent processes induced by an electromagnetic wave passage through the discharge gap.

The experiments were performed on a BIN generator with an output current amplitude of up to 270 kA and a rise time of 80 ns [3]. The impedance of the generator forming line was ~1 Ω, and the voltage at the generator output reached 240 kV, with the charging voltage of the forming line being of about 350 kV. The generator load was a rectangular ferrite ((Ni-Zn) $Fe_2O_4$) prism of grade M1000NN with transverse dimensions 10 x 20 $mm^2$. The prism was mounted perpendicular to the diode axis, see Fig. 1a. By changing the electrode length on the cathode side, the length of the active part of the ferrite prism was varied from 1.5 cm to 7 cm. The current flow path on the ferrite surface was set by a pattern drawn with a graphite pencil. This path was generally the same during consecutive discharges in the experiment [1,2]. The generator load was unmatched, and its impedance varied greatly during the pulse. The pressure in the discharge chamber did not exceed $10^{-4}$ Torr.

The radiation from the discharge was studied using calibrated diamond photoconductive detectors (PCDs) with flat spectral response C=5 ·$10^{-4}$ A/W in the energy range from 10 eV to 4 keV. In the high-energy region, the sensitivity smoothly decreases in accordance with the absorption of carbon [4]. The transverse size of the detector crystals was ~3x1 mm, whereas in the detection direction it was 0.5~mm, which provided sufficient sensitivity of the detector up to energies of 10 keV. Detector response time was less than 0.3 ns. The total time resolution of the recording channel was 2 ns, with the bandwidth of the Tektronix TDS 3104B oscilloscope and the cable lines taken into account. The detectors were typically placed at a distance of 20 cm from the end face of the ferrite prism at various angles with respect to the discharge direction (x axis in Fig. 1a).

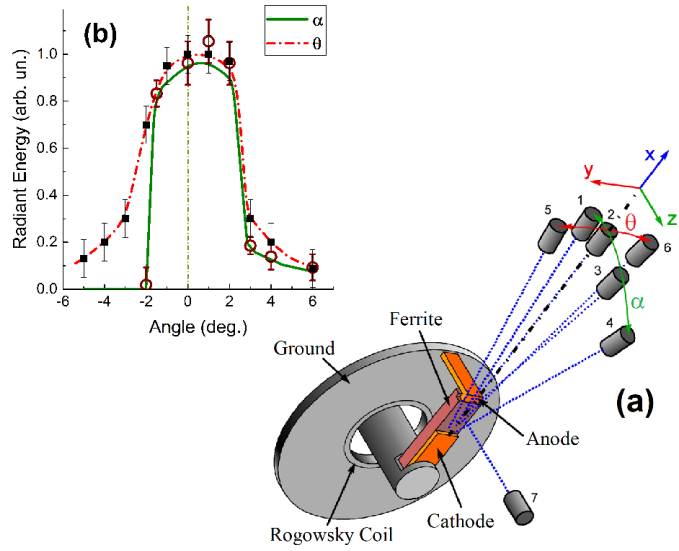

Fig. 1. (a) Experimental layout. (b) Angular intensity distribution for the radiation from ferrite.

The arrangement of the detectors is shown in Fig. 1a. The angular range of radiation detection in the ferrite surface plane xy was -8° < θ < +22° (azimuthal angle), and in the xz plane, orthogonal to the ferrite surface, it was -1° < α < +5 ° (polar angle). To estimate the width of energy distribution over the cross section of the generated beam, we used Fuji TR imaging plates sensitive to both x-ray and UV radiation; the plates were placed at a distance of 26 cm from the end face of the ferrite prism. In order to filter the bright UV-emission of the discharge, the films were covered by an aluminum foil. The time dependence of the load current was calculated by numerical integration of the signal from the Rogowski coil (with a bandwidth of >500 MHz). The output voltage on the discharge gap was measured by a resistive-capacitance voltage divider.

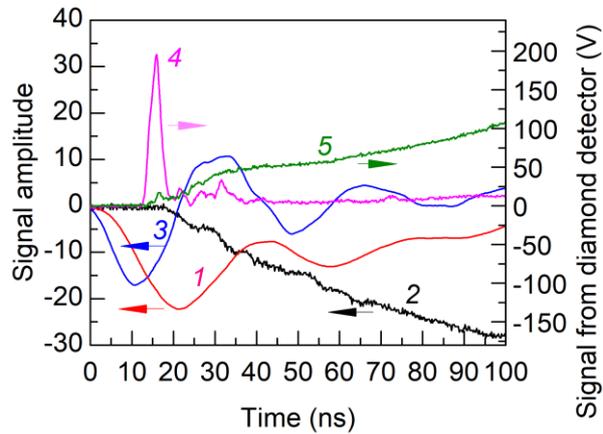

Fig. 2. The temporal dependencies of the applied voltage (1), discharge current (2), first derivative of the applied voltage (3), the radiation intensity observed along (4) and perpendicular (5) to the ferrite active surface (the positions 2 and 7 of the detectors in the Fig. 1a). The first detector was situated at $L = 26\ cm$ from the end of the prism, the second one was mounted at the $L_\perp = 26$ cm from the ferrite surface.

The time dependences of the discharge current, applied voltage and the radiation intensities recorded along and perpendicular to the ferrite surface (PCDs in positions 2 and 7) are shown in Fig. 2. Also, a first deviation of the voltage is presented in this figure. The experimental conditions were the following: the length of the active part of the ferrite prism was l = 6.5 cm; one detector was placed parallel to the active surface of the prism at a distance L = 26 cm from its end face, and the other was placed perpendicular to the active surface at the same distance.

It is seen in Fig. 2a that at the initial (prebreakdown) stage of the discharge, when there is almost no discharge current yet, a short ($\tau$ < 2 ns) radiation pulse is observed along the discharge axis, with the intensity of this pulse exceeding the intensities of radiation detected in the same time points in the perpendicular direction by an order of magnitude. Note that the actual signal duration is probably shorter since the measured value coincides with the time resolution of the recording channel.

Also, it should be noted, that the initial stage of the studied pulse formation coincides in time with a maximum of the first derivative of the applied voltage . In this point the speed of the voltage growth is maximal and at that time an inflection point in the dependence of the applied voltage on the time is reached. However, it is seen from the fig. 2 that the discharge current in this point is zero. After this time, the voltage is still increased and only when it reaches the maximum, the non-zero discharge current appears and a breakdown of the gap occurs.

We studied the angular distribution of radiation intensities. The detectors were placed at a distance L = 15 cm from the prism end face. The ferrite prism had a length l = 2 cm. The measurement results are presented in Fig. 1b. It can be seen that the radiation is concentrated in a region with angular sizes of ~4º and ~ 5 º (± 2.5º) in the planes perpendicular and parallel to the active surface of the ferrite prism. Regarding the fact that in our measurement geometry, the angular resolution is ~ 2º, the presented results should be thought of as an evaluation.

In vacuum discharges at such a high voltage (100…300 keV) high-energy electron beams can be formed [5,6]. Such a beam can cause a response on the registration equipment similar to the response caused by high-energy photons. For this reason, some experiments providing a proof of the electromagnetic origin of the studied radiation were carried out. First of all, a transmission of an aluminum filter with the thickness 50 μm and a beryllium filter with the thickness 10 μm (see below). Corresponding measurements allow to estimate an average energy of the electron beam [7]. In the case of the present work, for the Al filter ($I_{Al}$ = 0,09·$I_o$, where $I_o$ is an intensity without the filter) the energy of the beam should be $E_e$=95 keV when for the beryllium filter ($I_{Be}$=0,56·$I_o$) the beam energy should be $E_e$ = 55 keV. It is evident that these results contradict each other.

Another series of experiments was carried out when at the end of the ferrite prism was mounted a permanent magnet with the size *h=1 cm* providing a constant magnetic field with the intensity H = 700 G. An estimations show that a shift of the trajectory for electrons with the kinetic energy $E_e$=100 keV is Δy ≈ 20 cm. At the electron energy $E_e$= 50 keV this shift is Δy ≈ 32 cm. Our experiments show (fig. 1) that the deviation of the peak amplitude of the studied radiation does not exceed the 20% part of the peak average intensity. Magnetic field with the induction H = 700 G does not lead to measurable change in the amplitude of the radiation peak higher than this deviation. Consequently, it is proved that the nature of studied radiation is electromagnetic one.

A spectral composition of the radiation was investigated. For this purpose, integral intensities $I_{Al}$, $I_{Be}$ and $I_{PP}$ behind the aluminum, beryllium and polypropylene filters were measured. The filters had the following transmissions: *Al:* d = 50 μm, hv > 4 keV, (λ ≤2,5 Å), $I_{Al}$ = 0,09·$I_o$;

*Be*: d = 10 *μm*, h𝜈 > 0,5 *keV*, $I_{Be}$ = 0,56·$I_o$; *PP*(*C₃H₆* - polypropylene) d = 4 *μm*, 0,1 ≤ h𝜈 ≤ 0,293 *keV,* (124Å ≤ λ ≤ 42,4Å) $I_{PP}$ = 0,66·$I_o$]. Here the $I_o$ is an integral intensity measured in absence of the filters, i. e. in the region of quanta of the energies $10^{-2}$ keV ≤ h𝜈 ≤ 10 keV (1,24Å ≤ λ ≤ 1241Å).

The results of this measurement allow to estimate a spectral composition of the studied radiation using the database on the x-ray transmission spectra of different materials [8]. The results of this estimation are presented in Table 1. A dependencies of the energy spectral densities of the radiation on the photon energies in the region $10^{-2}$ keV≤h𝜈≤10 keV evaluated on the basis of the Table 1 data are presented in Fig 3.

Table 1. Energy fractions of the generated beam

| Energies of photons h𝜈 | Radiation intensity |
|---|---|
| *h* | *100%* |
| *1* | *10%* |
| keV ≤h𝜈≤0,293 keV, (42Å ≤ λ≤ 124Å) | |
| keV ≤h𝜈≤0,5 keV, (42Å ≤ λ≤25Å) | |
| keV≤h𝜈≤1 keV, (25Å ≤ λ≤12,4Å) | |
| keV≤h𝜈≤4 keV, (12,4Å ≤ λ≤3,1Å) | |
| keV ≤h𝜈≤10 keV, (3,1Å ≤ λ≤1,2Å) | |

*The measurement error (electrical noise and possible background radiation) is the same for all angles and does not exceed 10%*

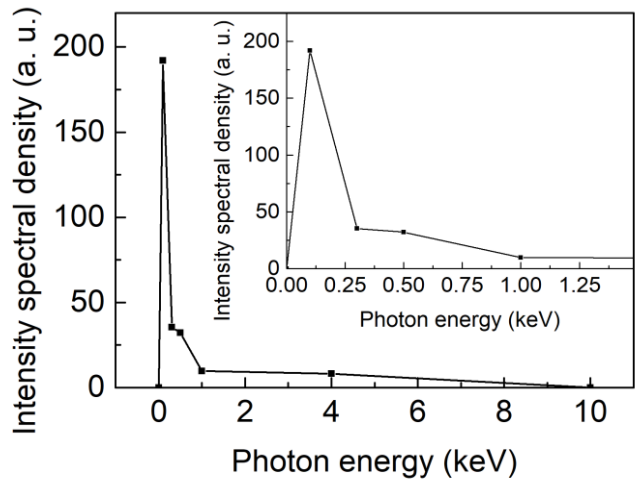

Fig. 3. The Intensity spectral density dependence on the energy of electromagnetic quanta for the studied radiation.

The Table 1 shows that the 50% part of the radiant energy corresponds to the x-ray radiation (>1 keV). Also, the most of the UV radiation has the energy higher than 1 eV. Let us note, that the discharge emission in the perpendicular direction to the ferrite surface lies in the region 10-800 eV [1,2].

Independent estimation of the angular distribution of the investigated radiation was done using the energy distribution over the cross section of the generated beam. Fuji TR imaging plates were used, placed at a distance L = 26 cm from the end face of the ferrite prism, with the length of its active part l = 4.5 cm.

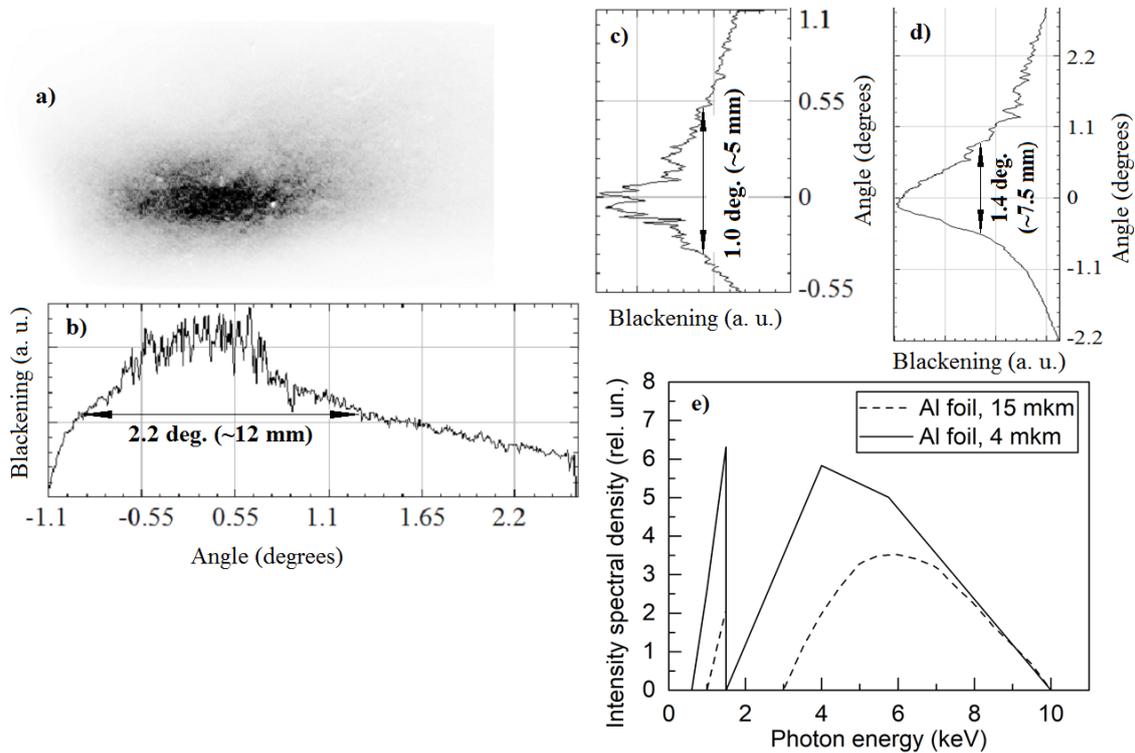

Fig. 4. (a) Image of the cross section of the generated x-ray beam obtained by Fuji BAS TR imaging plate coated with an aluminum filter (d= 15 μm) and positioned at the distance L = 26 cm from the end of the ferrite prism with the active length l = 4.5 cm; (b) The angular dependence of the plates darkening coated by the aluminum filter (d= 15 μm) along the image; (c) The angular dependence of the plates darkening coated by the aluminum filter (d= 15 μm) across the image; (e) The spectral densities of the studied radiation behind the aluminum filter: d = 4μm - solid curve and d = 15 μm – dashed curve.

The result is shown in Fig. 4. The darkening of the imaging plate corresponds to the time-integrated radiation energy emitted in the given direction. It is seen from the Fig 4, that the beam cross-section behind the aluminum filter (d=15 μm) is characterized by the angle size Δθ~2,2° (the corresponding linear size is $b$~ 0.5 cm). The cross-section longitudinal length of the beam behind the aluminum filter with the thickness d = 4 μm is nearly not changed while the transverse length is grown up to Δα~1,4 ° (Fig. 4 d). The corresponding linear length is $b$~ 0,75 cm. A comparison of energy spectral densities for radiation behind the aluminum filters with the widths d=4 μm and 15 μm (Fig. 4 e) shows that the radiation behind the filters at d=4 μm a part of low-energy quanta is much more than at d=15 μm.

The energy characteristics of the radiation were studied. The total energy registered by the detector is determined as follows

$$\varepsilon = \frac{1}{CR}\int_{-\infty}^{\infty} V(t)dt. \qquad (1).$$

Here, R = 75Ω is the detector load resistance, V (t) is the instantaneous value of the signal, measured in Volts, at time t, C=5·10⁻⁴A/Вт is the detector sensitivity. The duration of the radiation pulse may turn out to be significantly less than the time resolution of the recording channel (~2 ns), and the pulse duration and shape cannot be obtained by electrical measurements only. Nevertheless, since the signal spectral width (see Table I) is scarcely beyond the region of the detector spectral sensitivity (10 eV - 4 keV), relation (1) can be used to estimate the total energy incident on the detector. The value of $\int_{-\infty}^{\infty} V(t)dt$ was determined as the area under the curve V (t) (see Fig. 2a).

During measurements, the length of the ferrite active part *l* was varied from 1.5 cm to 7 cm by changing the length of the negative electrode. In several series of experiments, the length of the active part was first increased (2, 3, 5, 6, and 7 cm), and next decreased (6.5, 5.5, 4.5, 3.5, 2.5, and 1 cm). At each length in each series, 2 or 3 consecutive shots were made, with the total length of the ferrite sample being constant. For estimates, the beam cross section was considered a rectangle with dimensions s = 1 cm (width) and h = 0,2 cm (height). The value of h is smaller than the transverse dimensions on the detector crystals. The crystal sizes of the diamond detectors were $r_D$ ~ 0,3 cm in the lateral direction. This sizes were almost half as much than the smallest cross-sectional dimension of the beam measured behind the aluminum filter with the thickness d=15 *μm*. Thereby, the signal measured by the detector is proportional to the spatial energy flux *P* of the studied radiation.

$$P = \frac{\varepsilon}{S_D} \qquad (2).$$

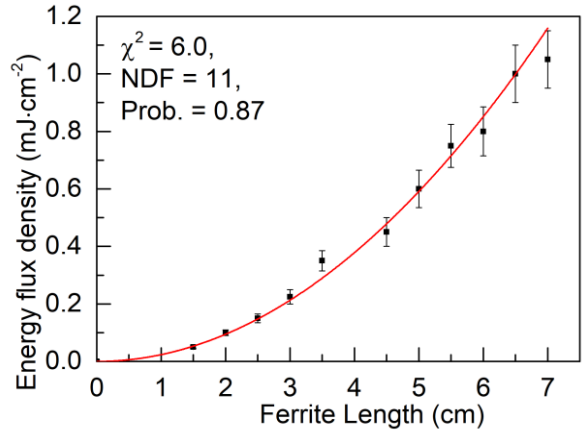

Fig. 5. The energy flux spatial density *P* dependence on the discharge gap length. The diamond detector was placed at the position 2 in Fig. 1a)

The energy flux dependence on the active length of the ferrite surface is depicted in Fig. 5 by squares. This figure shows that the energy flux P in nonlinearly-growing function of the ferrite surface active length. A solid line in the Fig. 5 shows a fit of the experimental data by a quadratic dependence. For the case under consideration, the pure quadratic dependence is statistically significant at the probability level W > 0.87. It was founded, that the maximal energy flux P at the discharge gap 7 cm was ~ 1 mJ·cm$^{-2}$. In order to estimate the total energy $\mathcal{E}$ of a single pulse of the studied radiation, we calculated a relation

$$\mathcal{E} = P \cdot S, \qquad (3),$$

where $S$ is the beam cross-section, which was regarded as a rectangle with the widths a = 1.2 cm and the height b = 0.5 cm (this size was founded to be the smallest behind the aluminum filter with the thickness d = 15μm). The maximal radiant energy of the radiation was ~0.6 mJ at the discharge gap 7 cm and the average radiant flux was 0.3 MW.

What is the physical nature of the observed radiation? In view of the sharp asymmetry in the angular distribution of the radiation intensity in the absence of focusing and limiting devices, one can conclude that the radiation is coherent, and the asymmetry in the spatial distribution is due to interference phenomena. We encounter a similar phenomenon when Cherenkov radiation arises, that is electromagnetic radiation of optically transparent media occurs, caused by a charged particle moving in a medium at a speed exceeding the speed of light in this medium [7--9]. The Cherenkov radiation condition can be derived considering the interference phenomena and Huygens-Fresnel principle.

What happens in our case? We assume that at the prebreakdown stage of the discharge, when a high voltage is applied to the cathode, a longitudinal electric field $E$ appears in the interelectrode gap (a bias current with the spatial density $J_{off} = \frac{1}{4\pi}\frac{\partial E}{\partial t}$ arises), which causes the magnetic field in the direction perpendicular to both $\boldsymbol{J}_{off}$ and to the normal $\boldsymbol{n}$ to the ferrite surface. Thus, a high-power magnetic field pulse passes through the discharge gap. This pulse induces short-lived magnetization of the ferrite surface. All the induced magnetic dipoles are oriented in the same direction with the forming magnetic field. In the geometry of the present experiment the symmetry of the system leads to the emission of interfering cylindrical elementary waves forming the resultant radiation.

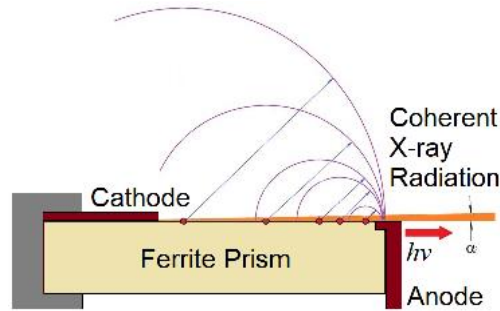

Fig. 6. Diagram explaining the formation of the coherent radiation region

Fig. 6 shows the diagram explaining the formation of the radiation as a result of coherent addition of elementary electromagnetic waves. The excitation pulse and the radiation it generates move in the same direction at the same speed. The envelope of the wavefronts of the elementary waves exists only in a small region near the ferrite surface where the phase matching of the radiation from elementary sources is ensured.

According to the Huygens-Fresnel principle, the elementary waves are mutually canceled except for their common envelope. Thereby, the emitted radiation has low angular divergence and propagates parallel to the ferrite surface towards the anode. The coherence of the radiation is, analogous to the Vavilov-Cherenkov radiation, due to the equivalent excitation conditions for all emitters. In a first approximation, one can be assumed that the longitudinal electric field strength $E$ is proportional to the applied voltage $U$. Thereby, a highly-directional electromagnetic pulse is formed when the bias current $J_{off} = \frac{1}{4\pi}\frac{\partial E}{\partial t}$ achieves its maximal value (see Fig. 2).

It can be shown that the energy flux P of the studied radiation in a case of the coherent interference of cylindrical unit waves can be calculated by a following expression:

$$P(y,l,L) \frac{(l)^2}{L}\left(SinC\left[\frac{\pi y^2 l}{2\lambda L^2}\right]\right)^2 \qquad (4).$$

In this formula $y$ is the spatial coordinate in the detector plane, $l$ is the length of the active part of the ferrite surface, $L$ is the distance between the end of the ferrite prism and the plane where the detectors are installed. The expression (4) was obtained using the approximation based on the relation $L \gg l,y$. The conclusion from (4) is that the energy flux $P$ is proportional to the second order of the active length of the ferrite surface if the radiation is coherent. This dependence was experimentally observed (Fig. 5).

In conclusion, we emphasize the important feature and novelty of the considered phenomenon: no optically transparent medium with a refractive index *n* and no charge moving at a speed *v* > *c/n* are involved. The studied radiation is formed as the electromagnetic excitation pulse passes over the surface of the ferrite prism. This pulse and the radiation it generates move in the same direction at the same speed, and the radiation region is formed as a result of coherent addition of elementary electromagnetic waves. Consequently, the radiation with high radiant intensity is emitted.

This study was supported by the Russian Science Foundation, project no. 19-79-30086.